\documentclass{aastex701}
\usepackage{wasysym}
\usepackage{xcolor}
\newcommand{\rev}[1]{#1}

\newcommand{\lsun}{L_{\odot}}

\newcommand{\rsun}{R_{\odot}}
\newcommand{\msun}{M_{\odot}}
\providecommand{\tablenotemark}[1]{\textsuperscript{#1}}

\shorttitle{Archival X-Ray Observations of TV~Cassiopeiae}
\shortauthors{Yang}
\submitjournal{PASP}

\begin{document}

\title{Archival X-Ray Observations of the Eclipsing Algol-type Binary TV~Cassiopeiae}

\author[orcid=0000-0002-7584-4756]{Jun Yang}
\email[show]{junyangpro@gmail.com}
\affiliation{Center for Astrophysics | Harvard \& Smithsonian, 60 Garden Street, Cambridge, MA 02138, USA}
\correspondingauthor{Jun Yang}

\begin{abstract}
TV~Cassiopeiae (TV~Cas) is an eclipsing Algol-type binary with a B9\,V primary and an evolved late-type secondary in a 1.81-day orbit. We examine six serendipitous \textit{Chandra} and two \textit{XMM-Newton} observations spanning 2003--2012. The supplied \textit{Chandra} spectra show thermal plasma near 1~keV and sub-solar fitted abundances, with an additional hotter component in one observation. EPIC-pn fits improve substantially with two temperatures: approximately 0.74/1.39~keV in 2003 and 0.87/2.16~keV in 2012. At 240~pc their nominal intrinsic 0.3--8~keV luminosities are $7.1\times10^{30}$ and $1.2\times10^{31}$~erg~s$^{-1}$, but residual structure, particularly in 2012, and response-calibration uncertainties limit physical interpretation. A prominent flare-like event occurs in \textit{Chandra} ObsID 08458 \rev{(2006-11-04)}. \rev{The longest observation carried out} in 2012 pn light curve spans approximately 82\% of an orbit and shows no conspicuous deep eclipse at the predicted secondary conjunction. An exploratory multi-epoch fold contains apparent troughs, but heterogeneous responses, intrinsic variability, and ephemeris uncertainty prevent a unique orbital interpretation. Thermal spectra and flare-like behavior favor a magnetically active secondary corona. The available products do not securely determine a coronal scale height, an orbital modulation amplitude, elemental abundance changes, or the fractional contributions of other unresolved emitters.
\end{abstract}


\section{Introduction} \label{sec:intro}

Algol systems are semi-detached binaries in which an evolved, active late-type star fills its Roche lobe and transfers matter toward an early-type companion \citep{sing95aas, sing95apj}. The X-ray interest in these systems stems from the expectation that the early-type star is X-ray dark: it is too hot to drive an outer convection zone and too cool to power a radiatively driven, shock-heated wind \citep{sch07}. \citet{chu04} demonstrated that the X-ray emission of Algol itself is dominated by the secondary. One possible exception would be Roche-lobe overflow from the cool star producing accretion onto the hot primary; no direct X-ray evidence for such accretion exists in Algol systems, although it has been inferred from H$\alpha$ \citep{Rich99}.

\citet{sing96} compared the X-ray emission of RS~CVn and Algol-type binaries and found no clear correlation between X-ray luminosity and Roche-lobe filling fraction across their Algol sample \citep{sing96ASPC}. \citet{drak03} investigated the coronal abundances of the Algol secondary, noting that HR~1099 and Algol show smooth light curves largely devoid of obvious flares during quiescence.

Eclipses are particularly powerful diagnostics: the changing occultation of emitting regions can identify their host star and constrain their spatial distribution, provided that the ephemeris, binary geometry, and intrinsic variability are sufficiently well characterized. \citet{whi90} argued, from the absence of an obvious eclipse in the harder of two EXOSAT bandpasses, that the hotter and cooler coronal plasma in AR~Lac occupy distinct regions. \citet{cul90} reached a similar conclusion from an EXOSAT observation of the RS~CVn system TY~Pyx, in which a hard component appeared un-eclipsed. These results are consistent with the radio measurements of \citet{owe77} and with the two-temperature (hard plus soft) decompositions of low-resolution active-star spectra by \citet{swa81}.

TV~Cas is located at $\mathrm{R.A.}=00^{\rm h}19^{\rm m}18\fs74$, $\mathrm{Dec.}=+59\arcdeg08\arcmin20\farcs56$. Its fundamental parameters, compiled from \citet{erd02} and \citet{kha92}, are listed in Table~\ref{para}. \citet{bak73} reported photoelectric light curves of TV~Cas and argued that the orbital period is not constant, which they interpreted as evidence for ongoing mass exchange between the components. \citet{lan83} presented four near-infrared light curves and, adopting the ephemeris of \citet{mar80}, $\mathrm{HJD}=2441595.3582 + 1.8125944\,E$, computed observed-minus-calculated (O$-$C) residuals. Their analysis places the secondary minimum at phase 0.502. Earlier studies \citep[e.g.,][]{fag99} describe TV~Cas as a B9\,V star paired with a larger, cooler, tidally distorted late-type companion that fills its Roche lobe. For comparison with rotation--activity relations, synchronous rotation would give $P_{\rm rot}=P_{\rm orb}=1.81$~d and $v_{\rm eq}\approx92$~km~s$^{-1}$ for $R_2=3.29\,\rsun$. We use this as a conditional assumption, not a measured rotation period. An initial account of the X-ray observations was presented by \citet{yang19proc}. Here we expand the analysis and, crucially, distinguish variability measured within individual observations from apparent structure produced by folding observations separated by years.

A third body has been proposed from long-term period variations \citep{erd02}. It is unresolved in the X-ray data. If it is a late-type star, its own corona could contribute to the measured emission, so this possibility is included in our interpretation (Section~\ref{sec:origin}).

\begin{deluxetable}{lclc}
\tablecaption{Fundamental parameters of TV~Cas\label{para}}
\tablehead{\colhead{Primary} & \colhead{Value} & \colhead{Secondary} & \colhead{Value}}
\startdata
Spectral type        & B9\,V   & Spectral type        & G5\,IV-III\tablenotemark{a} \\
$M\ (\msun)$         & 3.78    & $M\ (\msun)$         & 1.53  \\
$R\ (\rsun)$         & 3.15    & $R\ (\rsun)$         & 3.29  \\
$T_{\rm eff}$ (K)    & 10500   & $T_{\rm eff}$ (K)    & 5235  \\
$L\ (\lsun)$         & 107     & $L\ (\lsun)$         & 7     \\
\hline
Orbital period (d)   & \multicolumn{3}{l}{1.81260009} \\
Separation $a$ ($\rsun$) & \multicolumn{3}{l}{10.9\tablenotemark{b}} \\
Inclination (deg)    & \multicolumn{3}{l}{$\sim83$\tablenotemark{c}} \\
Distance (pc)        & \multicolumn{3}{l}{$239.7\pm1.2$ (Gaia EDR3 astrometry)\tablenotemark{d}} \\
\enddata
\tablenotetext{a}{Also classified F7\,IV by \citet{kha92}.}
\tablenotetext{b}{From Kepler's third law with the tabulated masses and period.}
\tablenotetext{c}{Adopted as a high (eclipsing) inclination; \citet{kha92}.}
\tablenotetext{d}{SIMBAD lists a Gaia EDR3 parallax of $4.172\pm0.0206$~mas for Gaia DR3 source 428246623544412416. Direct inversion gives $239.7\pm1.2$~pc; this formal uncertainty excludes parallax zero-point and unresolved-binary systematics. We use 240~pc as a reference distance. The Gaia DR3 release retains the EDR3 astrometry \citep{gaiadr3}.}
\tablecomments{Stellar parameters are compiled from \citet{erd02} and \citet{kha92}. R.A.\ $=00^{\rm h}19^{\rm m}18\fs74$, Dec.\ $=+59\arcdeg08\arcmin20\farcs56$.}
\end{deluxetable}

\section{Observations and Data Reduction} \label{sec:obs}

TV~Cas fell within the field of view of the \textit{Chandra X-ray Observatory} six times and of \textit{XMM-Newton} twice, in every case serendipitously during pointed observations of the starburst galaxy IC~10. The eight exposures span 2003--2012 and are summarized in Table~\ref{tab:obslog}. The \textit{Chandra} data were taken with ACIS-S; the \textit{XMM-Newton} data are from the EPIC-pn camera.

\begin{figure*}[ht!]
\plotone{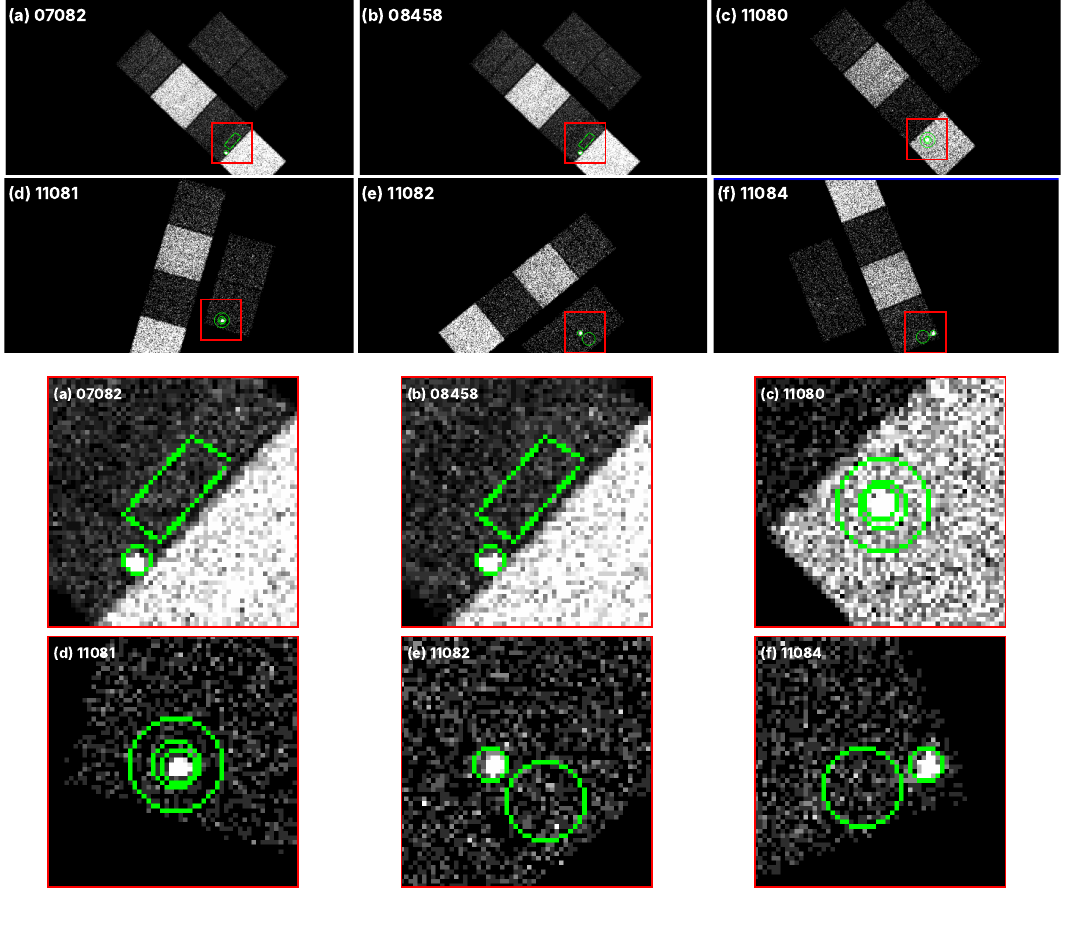}
\caption{Source and background extraction regions for TV~Cas in the six \textit{Chandra} observations. Top two rows: the full ACIS-S fields of ObsIDs 07082 (a), 08458 (b), 11080 (c), 11081 (d), 11082 (e) and 11084 (f), ordered left to right and top to bottom;  The red square in each panel marks the area enlarged in the corresponding cut-out in the lower part of the figure. TV~Cas is the bright compact source in every cut-out. For ObsIDs 11080 and 11081 the source aperture ($r=30\arcsec$) and the concentric background annulus  are shown in green; for ObsIDs 07082 and 08458 (green rectangle) and 11082 and 11084 (green circle) the green region marks the background area placed beside the source, and the $30\arcsec$ source circle is centered on the bright source. \label{fig:reg}}
\end{figure*}

The \textit{Chandra} light curves and spectra were extracted with \texttt{CIAO}~4.8 \citep{ciao}, with event reprocessing using \texttt{chandra\_repro} and barycentric correction using \texttt{axbary}. Because TV~Cas lies far off-axis and near detector boundaries, the source and background apertures were selected manually. Figure~\ref{fig:reg} shows their placement, including enlarged views. The source was extracted with an approximately $30\arcsec$ circular aperture. For ObsIDs 11080 and 11081 a concentric background annulus was used; the other observations used nearby rectangular or circular background regions. 

The \textit{XMM-Newton} EPIC-pn data were processed using the Science Analysis System \citep[\texttt{SAS};][]{sas}. The revised pn light curves are plotted directly from supplied numerical products, with their RATE and ERROR arrays in 200~s bins. For 2003, ObsID 0152260101, the corrected FITS product records 0.3--8~keV selection, barycentric TDB times with MJDREF 50814, fractional exposure, and background contribution. Its 219-bin grid spans 43.8~ks, whereas the event-header elapsed interval is 43.64~ks and the summed fractional-exposure equivalent is 37.21~ks; these are distinct measures of coverage. The user-identified 2012 ASCII product contains 646 TIME/RATE/ERROR rows spanning a nominal 129.2~ks bin grid.  
\begin{deluxetable*}{llcccccc}
\tablecaption{X-ray observations of TV~Cassiopeiae\label{tab:obslog}}
\tablehead{\colhead{ObsID/epoch} & \colhead{Instrument} & \colhead{Date (UT)} & \colhead{Start MJD} & \colhead{Exposure (ks)} & \colhead{Approx. phase} & \colhead{Mean rate (s$^{-1}$)} & \colhead{Description}}
\startdata
07082 & ACIS-S & 2006-11-02 & 54041.5 & 43.2 & 0.50--0.78 & 0.176 & two-temperature fit \\
08458 & ACIS-S & 2006-11-04 & 54044.0 & 44.2 & 0.87--0.15 & 0.183 & flare-like event \\
11080 & ACIS-S & 2009-11-05 & 55140.6 & 15.2 & 0.84--0.94 & 0.274 & highest ACIS rate \\
11081 & ACIS-S & 2009-12-25 & 55190.1 & 14.0 & 0.17--0.26 & 0.137 & rising rate \\
11082 & ACIS-S & 2010-02-11 & 55238.4 & 15.4 & 0.81--0.91 & 0.138 & variable appearance \\
11084 & ACIS-S & 2010-05-21 & 55337.8 & 14.8 & 0.65--0.74 & 0.075 & lowest ACIS rate \\
XMM 2003 & EPIC-pn & 2003-07-03 & 52823.8 & 43.8 & 0.67--0.95 & 0.598 & 219-bin grid; live equivalent 37.2 ks \\
XMM 2012 & EPIC-pn & 2012-08-18 & 56157.9 & 129.2 & 0.11--0.94 & 0.928 & longest exposure \\
\enddata
\tablecomments{ACIS dates, nominal exposures, and mean rates are from the data archive; their phase ranges derive from the reconstructed display. The phase interval 0.87--0.15 wraps through phase zero. The 2003 pn mean is fractional-exposure weighted from the FITS array. Its 43.8~ks bin grid includes partial exposure; its live equivalent is 37.21~ks. The original 2012 array has 646 bins, consistent with a nominal 129.2~ks grid. The pn phase ranges are approximate and do not establish live coverage. Rates are instrument-specific and are not corrected for differing off-axis responses. All pointings were directed at IC~10.}
\end{deluxetable*}

\section{Analysis and Results} \label{sec:lc}

We use the individual X-ray light curves to characterize variability on timescales from minutes to years, test whether the continuous 2012 observation contains the predicted eclipse of the cool component, and use the spectra to characterize the emitting plasma. A phase-folded view of all epochs is retained only as an exploratory display, and we show explicitly that its apparent structure is sensitive to differences between observations (Section~\ref{sec:lcmodel}).

\subsection{Individual light curves} \label{sec:fold}

\begin{figure*}[ht!]
\plotone{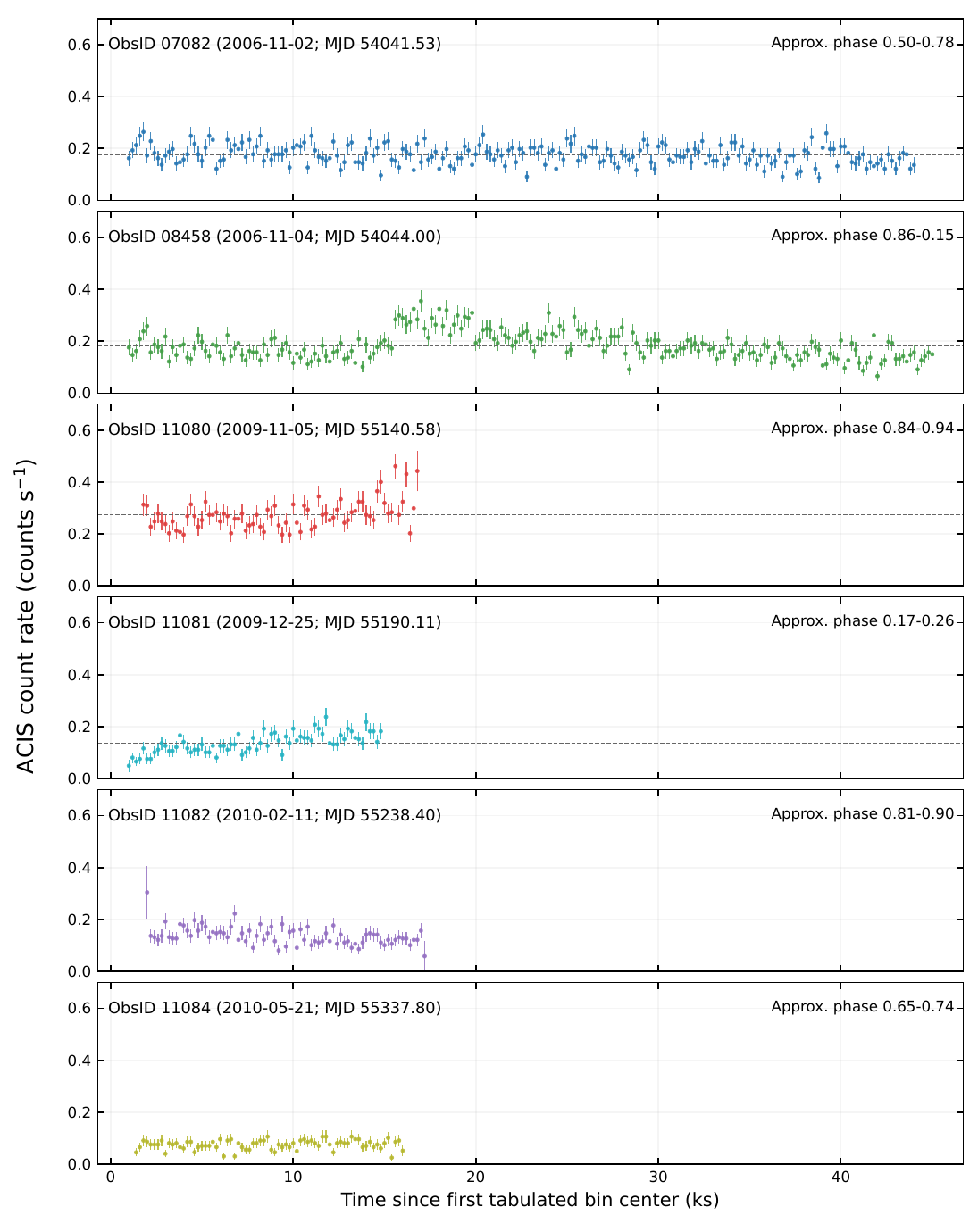}
\caption{\textit{Chandra} ACIS-S light curves of TV~Cas with numerical TIME, RATE, and ERROR arrays in 200~s bins. Panels show ObsIDs 07082, 08458, 11080, 11081, 11082, and 11084 on common time and count-rate scales. Error bars are the  measurement uncertainties; dashed lines show each observation's mean over the plotted bins. Rows with both zero rate and zero error are omitted; no positive-error rows are rejected on the basis of count rate. \rev{The date and MJD in each panel label give the start of the plotted light curve,
$\mathrm{MJD_{start}} = 50814 + t_1/86400$, where $t_1$ is the first tabulated bin
time in \textit{Chandra} mission seconds and MJD~50814 (1998 January 1) is the
standard \textit{Chandra} time reference (MJDREF). This reference is used only to
convert mission time to MJD. The phase ranges are computed separately from the
ephemeris of Equation~(1), with $P = 1.81260009$~d and
$T_0 = \mathrm{MJD}~32827.76$, and are approximate (Section~\ref{sec:fold}). Time is measured from the first tabulated bin center.} } 
\label{fig:eachlc}
\end{figure*}

\begin{figure*}[ht!]
\plotone{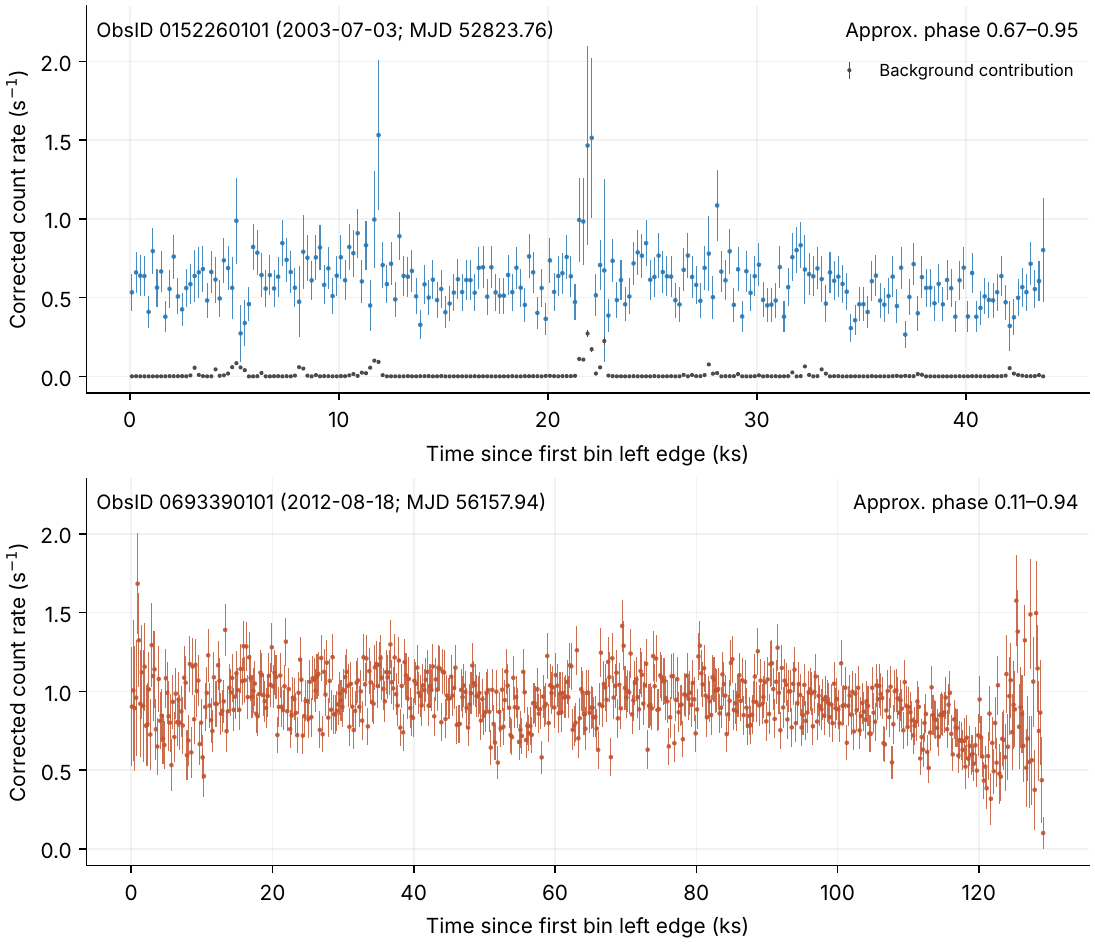}
\caption{\textit{XMM-Newton} EPIC-pn light curves \rev{of TV~Cas} in 200~s bins. Top: \rev{Top: ObsID 0152260101 (2003), corrected light curve}, with its background contribution (gray). Bottom: \rev{ObsID 0693390101 (2012)}. Both panels use a common count-rate scale; time is measured from the first bin's left edge. \rev{The labels give the start date and MJD of each plotted light curve and its approximate orbital-phase coverage, computed as for Figure~\ref{fig:eachlc}.} No points were discarded for low rate or low fractional exposure. The 2003 fractional exposures range from 0.12 to 0.91. }
\label{fig:xmmlc}
\end{figure*}

The six \textit{Chandra} light curves are shown together, on common time and count-rate scales, in Figure~\ref{fig:eachlc}, and the two EPIC-pn light curves in Figure~\ref{fig:xmmlc}. All are binned at 200~s. Orbital phases are computed with $P = 1.81260009$~d and phase zero at $T_0=\mathrm{MJD}~32827.76$,
\begin{equation}
\phi(t) = \left[\frac{t - T_{0}}{P}\right] \bmod 1 ,
\end{equation}
where $t$ is the barycentric time and ``$\bmod 1$'' denotes the fractional part. The adopted elements were used in the earlier proceedings analysis \citep{yang19proc}. Published O$-$C studies establish that a constant-period extrapolation is imperfect \citep{bak73,lan83,erd02}. In addition, the supplied graphical reconstruction applies a phase offset of 0.0027 to align the \textit{Chandra} points with the original folded figure. 

The light curves show variability within individual exposures. The most prominent flare-like event occurs in ObsID 08458 (the second panel of Figure~\ref{fig:eachlc}). Its rate approximately doubles near 16~ks after the start of the plotted interval and declines over roughly 10~ks. The rapid rise and slower decay are suggestive of a coronal flare. ObsID 11081 shows a sustained rise, and both EPIC-pn series show excursions. The 2003 background contribution rises during several short rate excursions; these bins also have reduced fractional exposure and larger propagated errors. This cautions against interpreting the excursions as stellar flares. 

The highest and lowest mean ACIS count rates differ by a factor of approximately 3.7. This is a measured contrast in the count-rate results, not a response-corrected luminosity ratio. Different off-axis responses, chip placement, aperture losses, and spectral shapes can contribute to the difference. With eight irregularly sampled epochs over nine years, the present products do not establish an activity cycle. 

\subsection{Phase folding and descriptive fits} \label{sec:lcmodel}

\begin{figure*}[ht!]
\plotone{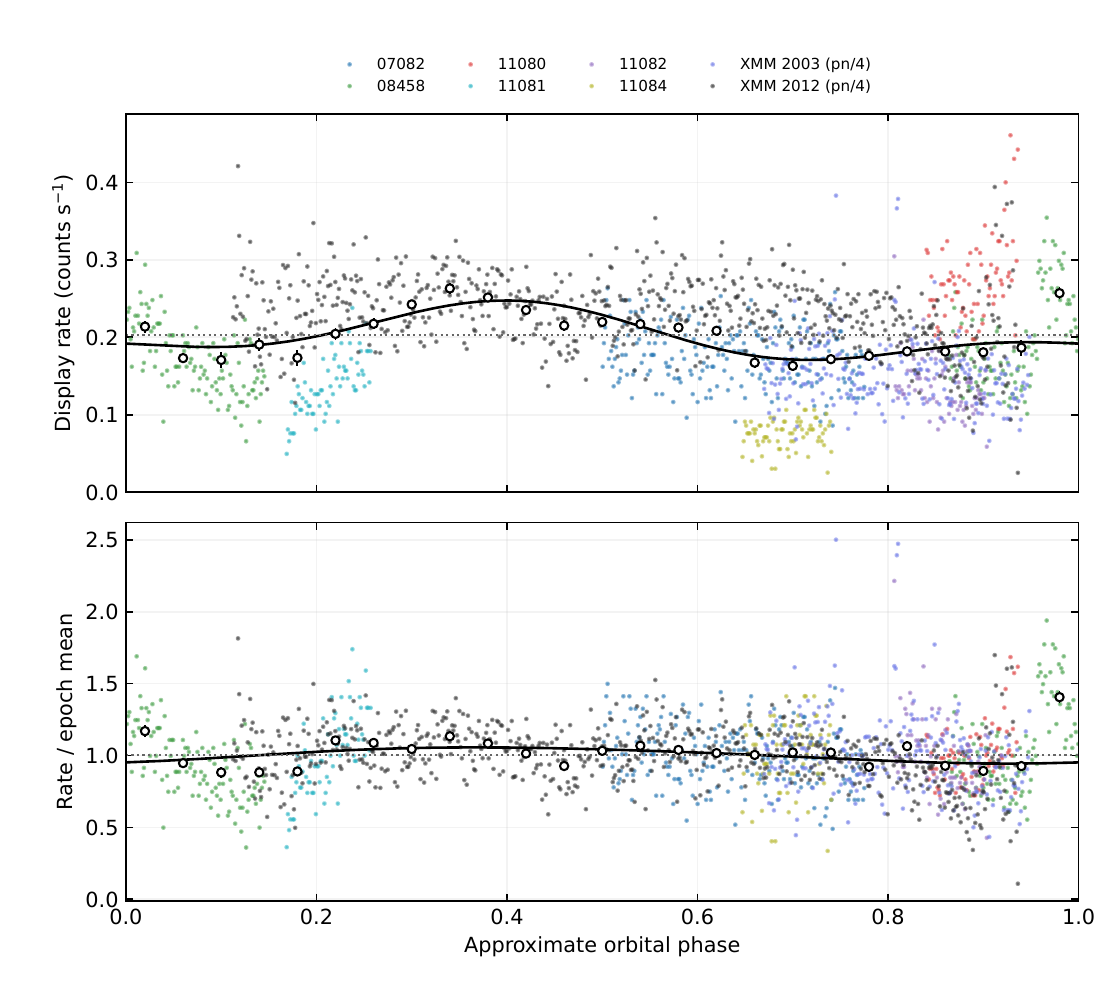}
\caption{Folded light curves for all six ACIS observations and both EPIC-pn epochs. Top: count rates, with pn divided by four for display only, without a flux conversion. Bottom: each observation divided by its own mean over the plotted bins. Colors distinguish observations. Open circles show 25 phase-bin means with standard errors of the rates within each phase bin. Solid curves are computed descriptive two-harmonic fits; dotted lines are fitted constants. The absolute phase convention is approximate. Instrumental differences, intrinsic variability, ephemeris uncertainty, and temporal correlations preclude treating the curves as a formal eclipse or orbital-modulation detection.\label{fig:foldlc}}
\end{figure*}

Figure~\ref{fig:foldlc} combines the original pn numerical arrays with the existing ACIS graphical reconstruction. EPIC-pn rates are divided by four for display only, without a flux calibration. The pn phases are calculated from the supplied time coordinates using MJDREF 50814 and the adopted ephemeris; this reference is recorded in the 2003 FITS header but assumed for the metadata-free 2012 ASCII file. The additional 0.0027 phase offset belongs only to the inherited ACIS reconstruction. Thus the absolute phase scale remains approximate.

We average the displayed rates in 25 phase bins and summarize their structure with a constant and a two-harmonic curve,
\begin{equation}
f(\phi)=a_0+\sum_{k=1}^{2}[a_k\cos(2\pi k\phi)+b_k\sin(2\pi k\phi)].
\end{equation}
The descriptive weighted residual sums in Table~\ref{tab:lcfit} use standard errors of the rates in each bin. They omit temporal correlations, ephemeris uncertainty, and instrumental calibration differences and cannot be interpreted as calibrated goodness-of-fit probabilities or eclipse significances. We do not apply an ordinary Gaussian-trough $F$-test.

After dividing each observation by its arithmetic mean, the harmonic peak-to-trough amplitude decreases from approximately 44\% to 8\%. However, appreciable residual structure remains. Normalization can also remove genuine modulation that changes little during a short pointing. The result demonstrates sensitivity to epoch-dependent levels and heterogeneous responses; it does not establish the absence of orbital modulation or uniquely identify the cause of the apparent troughs.

\begin{deluxetable}{lcc}
\tablecaption{ Descriptive folded diagnostics\label{tab:lcfit}}
\tablehead{
\colhead{Quantity} &
\colhead{Display rates} &
\colhead{Per-epoch normalization}
}
\startdata
Phase bins & 25 & 25 \\
Constant residual sum/dof & 573.7/24 & 317.5/24 \\
Harmonic residual sum/dof & 156.2/20 & 252.3/20 \\
Harmonic peak-to-trough / mean & 0.379 & 0.113 \\
\enddata
\tablecomments{
Computed using the numerical light curves for all six
ACIS observations and both EPIC-pn epochs. For the display-rate
column, pn rates are divided by four for visualization only,
without a flux conversion. For the normalized column, each
observation is divided by its own mean over the plotted bins.
The residual sum is $\sum_i[(\bar r_i-f_i)/{\rm SEM}_i]^2$,
where ${\rm SEM}_i$ is the standard error estimated from the
scatter of rates within each phase bin. The constant model has
one parameter and the two-harmonic model has five. Degrees of freedom are the number of phase bins minus the number
of fitted parameters. The residual sums are descriptive measures
of the folded profile; temporal correlations, ephemeris uncertainty,
and instrumental differences are not included. They therefore do
not provide calibrated goodness-of-fit probabilities or establish
the statistical significance of orbital modulation.
}
\end{deluxetable}

\subsection{Eclipse test with the 2012 observation} \label{sec:geom}

Continuous eclipse observations can constrain coronal structure in active binaries \citep[e.g.,][]{whi90,drak14}. The 2012 EPIC-pn exposure is the only observation here spanning most of a single orbit. Its nominal duration is $129.2$~ks, approximately 0.82 of the adopted period. Using the assumed mission time reference, the original 2012 bin centers cover phases approximately 0.113--0.936. This is a phase assignment, not evidence of uninterrupted live exposure.

For a circular orbit and the adopted illustrative geometry ($a=10.9\,\rsun$, $i=83\arcdeg$, $R_1=3.15\,\rsun$, $R_2=3.29\,\rsun$), the stellar discs overlap near secondary conjunction over approximately $|\phi-0.5|<0.10$. The 2012 light curve does not show a conspicuous, isolated deep trough at this conjunction. Intrinsic variability, uncertainties in the background contribution,
and uncertainty in the extrapolated orbital phase prevent a robust quantitative upper limit on the eclipse depth from being established in the present analysis.

The absence of an obvious deep eclipse is compatible with extended coronal structures, emission concentrated outside the occulted region, emission from another unresolved component, or variability that masks an eclipse. It does not yield a unique coronal radius or scale height. An exponential-corona calculation would additionally assume a spatial emissivity law and a stable out-of-eclipse baseline; the existing temporal products alone cannot establish these assumptions or a confidence bound. New full-orbit observations with a contemporaneous ephemeris would permit a more reliable separation of eclipse geometry from coronal variability.

\subsection{X-ray spectra} \label{sec:spec}

\begin{figure*}[ht!]
\plotone{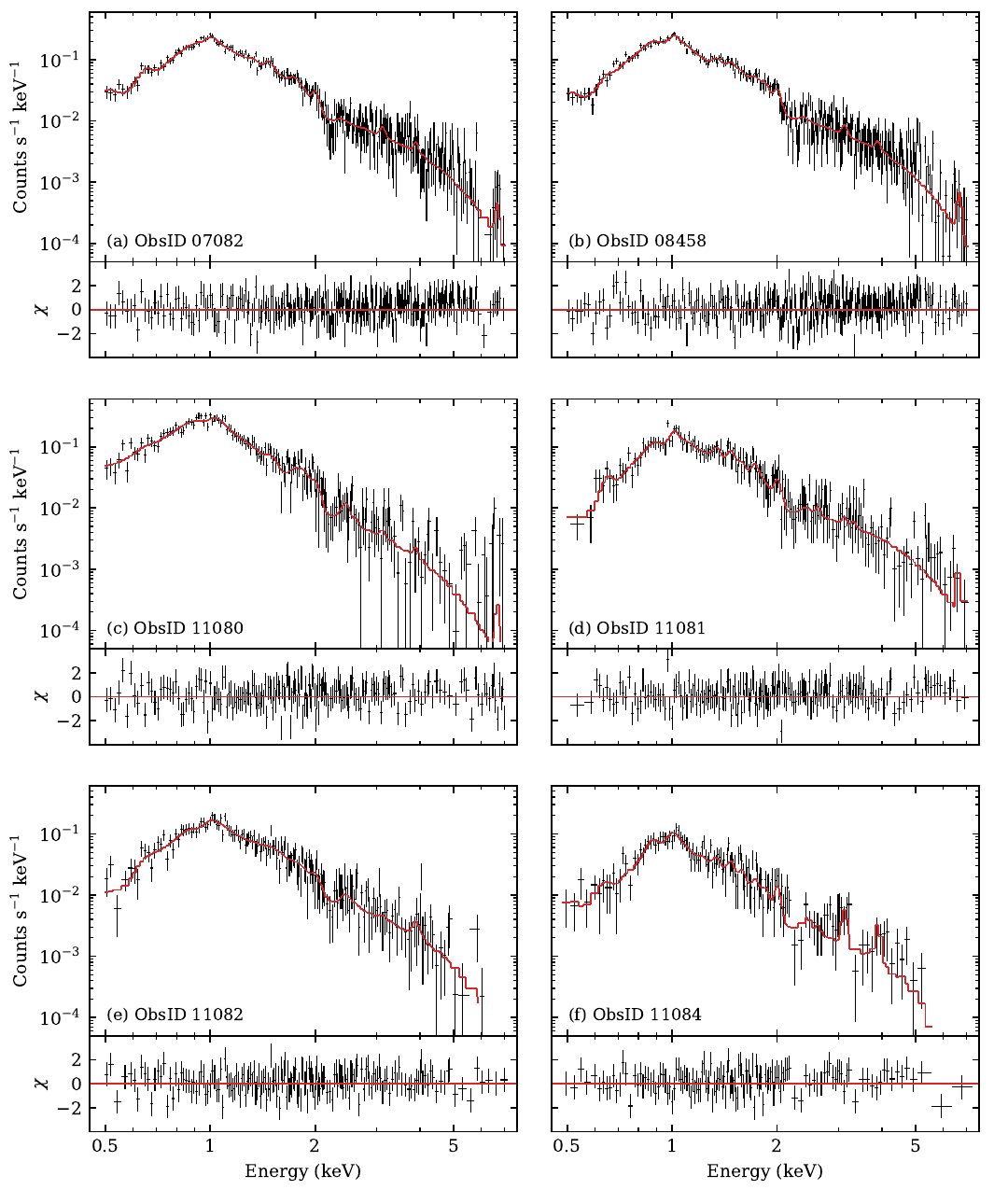}
\caption{\textit{Chandra} ACIS-S count spectra of TV~Cas for ObsIDs 07082,
08458, 11080, 11081, 11082, and 11084 (panels a--f), shown on common axes.
\rev{Red curves are the fitted models. Lower panels show the residuals
$\chi_i = (D_i - M_i)/\sigma_i$, where $D_i$ is the observed count rate in
spectral bin $i$, $M_i$ is the model-predicted count rate in that bin, and
$\sigma_i$ is the $1\sigma$ uncertainty on $D_i$; the error bars on $\chi_i$
are therefore $\pm1$. The panels were redrawn from the vector coordinates of
the original \texttt{ISIS} plots on common axis ranges.}%
 \label{fig:spec}}
\end{figure*}

Figure~\ref{fig:spec} shows the count-spectra of the six \textit{Chandra} observations. Table~\ref{model} includes the single-abundance fit values; the additional VAPEC fits are described separately in Section~\ref{sec:newfits}. These were obtained in \texttt{XSPEC} \citep{arnaud96} with Wisconsin photoelectric absorption \citep{mor83} multiplying an APEC optically thin thermal plasma \citep{smi01}. The single-abundance fits do not identify the solar abundance table, grouping rule, degrees of freedom, or confidence convention for all fitted parameters; this paper explicitly identifies the absorption upper limits as $1\sigma$. We therefore do not relabel these results as new \texttt{aspl} fits or 90\% confidence intervals.

Five observations were described by one temperature near 1~keV. ObsID 07082 was fitted by two components near 1.04 and 2.55~keV. New response-folded pn fits are presented in Section~\ref{sec:pn_spec}, with their model-adequacy limitations.

For reference, we convert the tabulated APEC normalizations into volume emission measures at $d=240$~pc using
\begin{equation}
\mathrm{EM}=\int n_e n_{\rm H}\,dV=10^{14}\,4\pi d^2\mathcal{N}.
\end{equation}
These are algebraic conversions of the original normalizations, not newly fitted quantities. Their systematic uncertainty includes the spectral model and distance; an emitting density additionally requires an emitting volume. 

\begin{deluxetable*}{lccccc}
\tablecaption{Original \textit{Chandra} spectral fits and derived emission measures\label{model}}
\tablehead{\colhead{ObsID} & \colhead{$N_{\rm H}$ ($10^{22}$ cm$^{-2}$)} & \colhead{$kT$ (keV)} & \colhead{$Z/Z_\odot$} & \colhead{EM ($10^{53}$ cm$^{-3}$)} & \colhead{Reduced $\chi^2$}}
\startdata
07082 & $<0.021$ & $2.55^{+0.17}_{-0.15}$ & $0.27^{+0.06}_{-0.06}$ & 6.20 & 315.7/268 \\
      &          & $1.04^{+0.02}_{-0.02}$ & & 4.14 & \\
08458 & $0.107^{+0.013}_{-0.012}$ & $1.059^{+0.017}_{-0.017}$ & $0.068^{+0.006}_{-0.006}$ & 20.7 & 348.2/281 \\
11080 & $0.106^{+0.018}_{-0.017}$ & $1.042^{+0.023}_{-0.024}$ & $0.103^{+0.014}_{-0.013}$ & 14.5 & 143.1/188 \\
11081 & $0.095^{+0.031}_{-0.028}$ & $1.285^{+0.051}_{-0.058}$ & $0.096^{+0.025}_{-0.022}$ & 15.2 & 123.8/180 \\
11082 & $<0.039$ & $1.268^{+0.053}_{-0.063}$ & $0.102^{+0.029}_{-0.025}$ & 11.7 & 101.1/174 \\
11084 & $<0.097$ & $1.081^{+0.217}_{-0.055}$ & $0.087^{+0.026}_{-0.021}$ & 8.96 & 72.7/131 \\
\enddata
\tablecomments{Values and fitted-parameter errors are carried over from the submitted Wisconsin-absorption/APEC fits. The original absorption upper limits are $1\sigma$; the remaining error convention is not documented in the supplied fit records and must not be interpreted as a verified 90\% convention. The abundance scale is that of the original fits, whose solar table is not recorded. EM is derived from the original normalizations at 240~pc, with no fitted uncertainties added. For 07082 the two EM values correspond to the temperatures on their respective rows. Response-integrated fluxes and EPIC-pn fit products are unavailable in the supplied materials. $\chi^2$/dof values are reconstructed from the original reduced $\chi^2$,
assuming dof equals the number of spectral bins minus the number of free
parameters.}
\end{deluxetable*}

\subsection{Additional Chandra fits} \label{sec:newfits}
Addtionally we also used ISIS\footnote{\url{https://space.mit.edu/cxc/isis/}} with \texttt{tbabs}$\times$(\texttt{vapec}$_1$+\texttt{vapec}$_2$), the Wilms abundance table (\texttt{wilm}; \citealt{wil00}), and a fitted interval of 0.5--7~keV. Their \texttt{rebin\_data(all\_data,3)} command requests a minimum of three counts per bin, not a three-channel grouping or a 20-count threshold. It used Verner baseline cross-sections of the absorption routine.

These highly flexible fits are not adopted as validated replacements for Table~\ref{model}. The 11080 shows explicitly failed minimization. In 11081 the fitted $N_{\rm H}$ reaches the imposed upper bound of $0.097\times10^{22}$~cm$^{-2}$; this boundary is not an observational upper limit. All three contain abundance parameters on imposed boundaries. The fitting allow component-specific elemental abundances, with some allowed maxima far above solar, and the limited counts per bin make a Gaussian $\chi^2$ interpretation questionable. A small reduced $\chi^2$ alone therefore does not establish convergence, reliable abundances, or the necessity of a second temperature. Uniform, more parsimonious models, suitable count statistics, and confidence searches are needed before a physical comparison.

\subsection{XMM EPIC-pn spectral fits}\label{sec:pn_spec}

We retrieved the pipeline source and background spectra and source-specific ARFs for ObsIDs 0152260101 (2003) and 0693390101 (2012) from the \textit{XMM-Newton} Science Archive. The matched source is within $1.84\arcsec$ and $0.74\arcsec$, respectively, of the SIMBAD position of TV~Cas. These pipeline apertures have radius approximately $40\arcsec$, larger than the approximately $15\arcsec$ source aperture in the manual 2003 light curver extraction. Their spectral exposures are 36.621 and 113.704~ks; these are distinct from the light-curve bin-grid spans. We retain the pipeline source/background EXPOSURE and BACKSCAL values. No new event screening was performed. The manually extracted 2003 histogram remains in the accompanying analysis package as an independent count-total check, but is not the input to the fits presented here.

The EPIC-pn pipeline products were generated with SAS~21.0.0.
The source-specific ancillary response files (ARFs) were paired
with the official full-frame, single-plus-double-event
redistribution matrices
\texttt{epn\_e1\_ff20\_sdY6\_v22.0.rmf} and
\texttt{epn\_e2\_ff20\_sdY7\_v22.0.rmf}
for the 2003 and 2012 observations, respectively. The matrix
selections follow the observing epoch and detector-position
identifiers recorded in the spectral headers. The analysis uses
archived ARFs and version~22.0 redistribution matrices, without
additional event reprocessing. Systematic uncertainties associated
with this calibration combination and the off-axis extraction
are not included in the quoted statistical intervals.

We fit 0.3--8~keV in XSPEC~13.0.1 (PyXspec~2.2.0), using \texttt{tbabs}$\times$\texttt{apec} and \texttt{tbabs}$\times$(\texttt{apec}+\texttt{apec}). The APEC emissivities use AtomDB~3.1.3, and the solar abundance table is \texttt{aspl} (\citealp{asplund09}), for both absorption and emission. TBabs version~2 uses its built-in Verner-based cross sections. Redshift is fixed to zero and the two thermal components share one fitted metal scale. The spectra were fitted over 0.3--8~keV using the pipeline
grouping, with bad-quality channels excluded. We used XSPEC's
W statistic, obtained by selecting \texttt{cstat} with a
Poisson background spectrum. Fits initialized at several
different temperatures converged to the same two-temperature
minimum. Quoted uncertainties are one-parameter 90\% confidence
intervals obtained from profile searches with $\Delta W=2.706$,
allowing the other free parameters to vary.%

\begin{figure*}[ht!]
\plotone{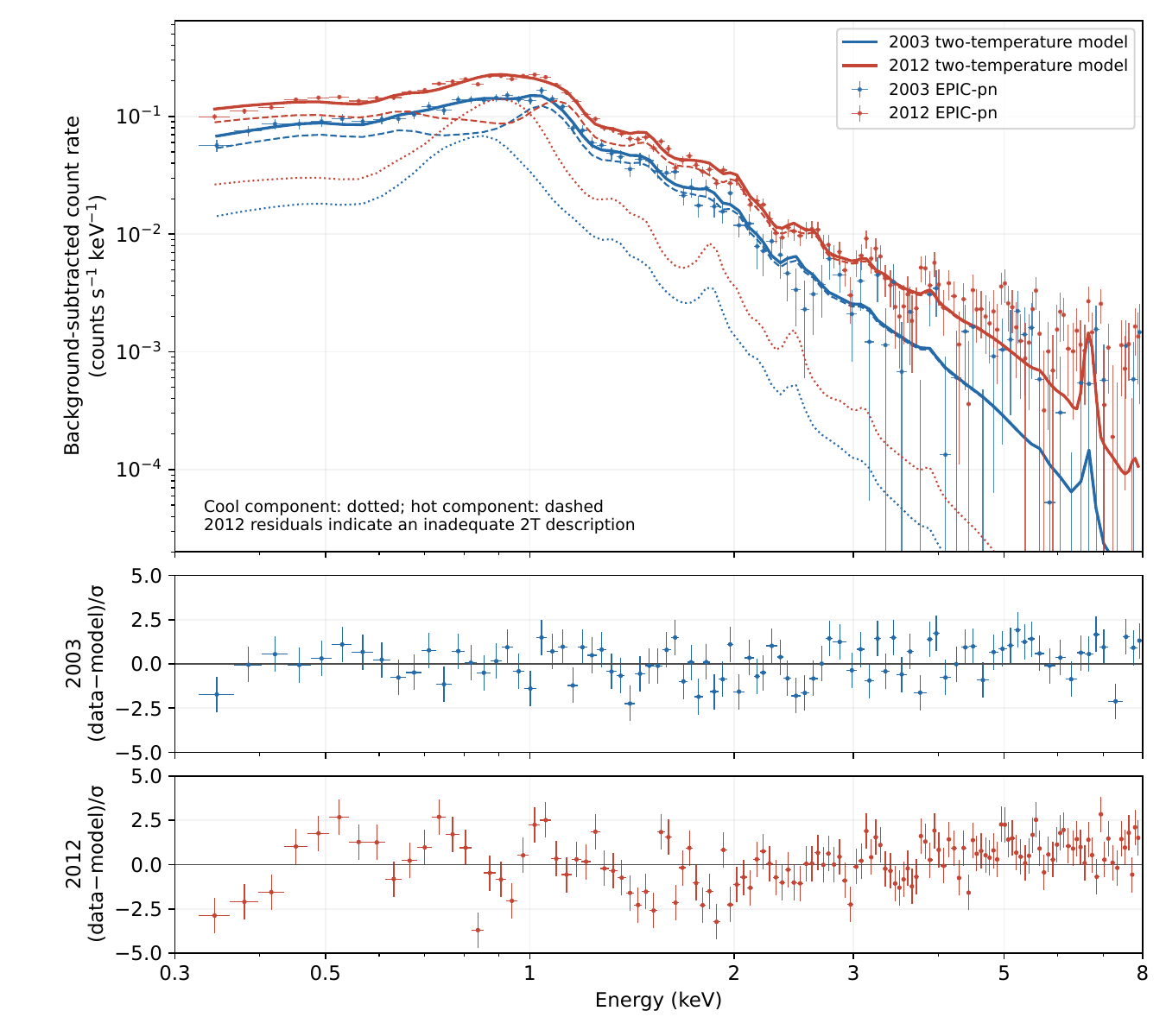}
\caption{Both EPIC-pn spectra and new two-temperature fits on common axes: 2003 (blue) and 2012 (red). Points are background-subtracted XSPEC count-rate displays, with pipeline grouping, horizontal bin widths, and XSPEC display errors. Solid curves are response-folded total models; dotted and dashed curves are the cooler and hotter components. Lower panels show display residuals divided by the display errors; the fit itself uses W, not these Gaussian residuals. The 2012 residual structure limits physical interpretation of the fitted parameters. These spectra use pipeline apertures and are distinct from the earlier manual 2003 histogram.\label{fig:pn_spec}}
\end{figure*}

Table~\ref{tab:pn_fits} gives the parameters and statistics, and Figure~\ref{fig:pn_spec} compares both epochs with their response-folded models, separate thermal components, and residuals. Adding a second temperature lowers W by 32.22 (2003) and 224.05 (2012). These improvements are not assigned a formal additional-component significance: the null model has a boundary normalization and an unidentified temperature. The fitted temperatures are approximately 0.74 and 1.39~keV in 2003, and 0.87 and 2.16~keV in 2012.

Goodness-of-fit simulations indicate residual structure beyond
that expected from counting statistics, particularly in 2012.
In 200 simulated spectra per observation, each refitted with the
two-temperature model, 95\% and 99\% had Pearson test statistics
below the observed values for 2003 and 2012, respectively.
The quoted parameter intervals are therefore conditional on the
adopted model and exclude uncertainties associated with model
inadequacy and calibration.
 These small simulation ensembles provide a diagnostic rather than a precise tail probability, and omit calibration uncertainty. A 0.5--7~keV check gives W/dof of 81.63/72 and 188.91/111; the 2012 absorbing column reaches zero and the abundance rises to approximately 1.04 solar. Merging pipeline groups until they contain at least 20 source and five background counts leaves the fitted 0.3--8~keV results unchanged. A three-temperature 2012 trial has competing minima; its lowest W solution drives the hottest component to the imposed 40~keV limit. We do not adopt that boundary-valued component as a coronal measurement. The apparent inter-epoch differences in absorption and abundance
are sensitive to background treatment, response calibration, and
the assumed plasma temperature structure. They therefore do not
establish intrinsic changes in these quantities. The quoted
90\% confidence intervals are conditional on the adopted
two-temperature model and exclude these systematic uncertainties.

Absorbed and intrinsic fluxes in Table~\ref{tab:pn_flux} are model integrations over 0.3--8~keV. Their 90\% intervals use \texttt{cflux} profile searches with a fixed reference normalization and a free relative component normalization; absorption, temperatures, and shared abundance are refitted. For the intrinsic flux, \texttt{cflux} is placed inside the absorption component. At 240~pc, the nominal intrinsic luminosities are $7.06\times10^{30}$ and $1.17\times10^{31}$~erg~s$^{-1}$, a ratio of 1.66. The table propagates flux intervals at fixed distance; the formal $1.2$~pc distance error contributes an additional approximately 1\% luminosity uncertainty, separate from model and calibration systematics. The quoted uncertainties are conditional on the adopted spectral
model and exclude calibration and model-related systematic effects. Emission measures use $\mathrm{EM}=10^{14}4\pi d^2\mathcal{N}$ and are conditional on the adopted thermal model.

\begin{deluxetable*}{cccccccc}
\tablecaption{Two-temperature EPIC-pn fits\label{tab:pn_fits}}
\tablewidth{0pt}\tabletypesize{\scriptsize}
\tablehead{\colhead{Epoch} & \colhead{$N_{\rm H}$} & \colhead{$kT_1$} & \colhead{$kT_2$} & \colhead{$Z/Z_{\odot}$} & \colhead{$\mathcal{N}_{1},\mathcal{N}_{2}$} & \colhead{1T W/dof} & \colhead{2T W/dof} \\ & \colhead{$10^{21}$ cm$^{-2}$} & \colhead{keV} & \colhead{keV} & \colhead{\texttt{aspl}} & \colhead{$10^{-4}$ cm$^{-5}$} & &}
\startdata
2003 & $0.77^{+0.22}_{-0.22}$ & $0.74^{+0.09}_{-0.11}$ & $1.39^{+0.17}_{-0.12}$ & $0.33^{+0.17}_{-0.11}$ & 2.00, 8.70 & 129.58/84 & 97.36/82 \\
2012 & $0.41^{+0.09}_{-0.11}$ & $0.87^{+0.02}_{-0.03}$ & $2.16^{+0.23}_{-0.16}$ & $0.74^{+0.21}_{-0.14}$ & 2.29, 10.46 & 475.44/128 & 251.39/126 \\
\enddata
\tablecomments{Errors are one-parameter 90\% profile intervals, $\Delta W=2.706$. The 1T and 2T models have four and six free parameters and 88 and 132 noticed bins in 2003 and 2012. Z is linked between the components. The 2012 fit retains substantial residual structure; parameters and errors are model-conditional.}
\end{deluxetable*}

\begin{deluxetable*}{ccccc}
\tablecaption{Model-conditional EPIC-pn fluxes, luminosities, and emission measures\label{tab:pn_flux}}
\tablewidth{0pt}\tabletypesize{\scriptsize}
\tablehead{\colhead{Epoch} & \colhead{$F_{\rm abs}$} & \colhead{$F_{\rm unabs}$} & \colhead{$L_{\rm unabs}$} & \colhead{$\mathrm{EM}_{1},\mathrm{EM}_{2}$} \\ & \colhead{$10^{-12}$ erg cm$^{-2}$ s$^{-1}$} & \colhead{$10^{-12}$ erg cm$^{-2}$ s$^{-1}$} & \colhead{$10^{30}$ erg s$^{-1}$} & \colhead{$10^{53}$ cm$^{-3}$}}
\startdata
2003 & 0.826 (0.798--0.858) & 1.024 (0.965--1.089) & 7.058 (6.651--7.503) & 1.38, 6.00 \\
2012 & 1.540 (1.503--1.592) & 1.700 (1.662--1.743) & 11.719 (11.453--12.013) & 1.58, 7.21 \\
\enddata
\tablecomments{0.3--8~keV values from the 2T model. Parentheses contain 90\% flux-profile ranges. Luminosity intervals assume fixed 240~pc distance; add its formal approximately 1\% uncertainty separately. Emission measures are listed without statistical uncertainties.
The quoted intervals exclude calibration uncertainties and
systematic effects associated with model inadequacy, which are
particularly relevant to the 2012 spectrum.}
\end{deluxetable*}

\section{Discussion} \label{sec:disc}

\subsection{Origin of the X-ray emission} \label{sec:origin}

The X-ray apertures do not resolve the two stellar components or the proposed tertiary. Nevertheless, the observed characteristics favor a magnetic corona on the late-type secondary. The B9\,V primary lacks the substantial surface convection needed for a solar-like dynamo and does not have the powerful wind of an early O/B star. Studies of X-ray-associated A-type stars also emphasize the role of late-type companions \citep{sch07}; they do not supply a universal numerical upper bound for this particular B9\,V star.

The evolved cool secondary is a more natural coronal host. Its thermal spectrum near 1~keV and the flare-like event in ObsID 08458 resemble magnetically active binary stars \citep{sing96,swa81,pan12}. The identification of the cool star as the X-ray host is also consistent with the spatial and spectroscopic evidence for Algol itself \citep{chu04}. This analogy is an interpretation rather than a localization measurement for TV~Cas.

Mass transfer makes accretion-related emission physically possible \citep{Rich99}. The thermal spectra and flare-like variability favor a magnetic
corona on the late-type secondary. Contributions from accretion
or an unresolved tertiary remain possible, but their relative
contributions are not constrained by the present data.

\subsection{Comparison with other Algols and active binaries} \label{sec:compare}

The characteristic temperatures and flare-like behavior are qualitatively consistent with the active Algol and RS~CVn systems studied by \citet{sing96} and \citet{swa81}. A quantitative luminosity comparison requires fluxes integrated over a stated energy band, a distance with uncertainties, and compatible bandpasses across the comparison sample. The new pn fits give model-conditional 0.3--8~keV luminosities of $7.06\times10^{30}$ and $1.17\times10^{31}$~erg~s$^{-1}$. A detailed comparison with other systems still requires compatible bandpasses and adequate spectral models.

As a conditional benchmark, synchronous rotation gives $P_{\rm rot}=1.81$~d for the secondary. Such rapid rotation can be associated with saturated magnetic activity, whose commonly used scale is $L_{\rm X}/L_{\rm bol}\sim10^{-3}$ \citep{piz03,wri11}. For the tabulated $L_{\rm bol,2}=7\,\lsun$, this corresponds to approximately $2.7\times10^{31}$~erg~s$^{-1}$. This number is a predicted comparison scale, not a measured luminosity of TV~Cas. The rotation--activity calibrations are primarily based on other stellar populations; application to a Roche-lobe-filling evolved secondary is indicative. If all fitted intrinsic pn emission is assigned to the secondary, the nominal 0.3--8~keV ratios are $L_{\rm X}/L_{\rm bol,2}\approx2.6\times10^{-4}$ and $4.4\times10^{-4}$. These are band- and model-dependent, especially in 2012, and do not establish a precise saturation level. %

\subsection{Coronal abundances} \label{sec:abund}

The new pn 2T models give shared metal scales of approximately 0.33 and 0.74 on the \texttt{aspl} scale, but the 2012 residuals and bandpass sensitivity prevent a secure inter-epoch abundance inference. The original fits give a globally scaled metal abundance of approximately 0.07--0.27 on their adopted solar scale (Table~\ref{model}). The additional VAPEC tests (Section~\ref{sec:newfits}) fit individual elements independently, but their boundary behavior and missing uncertainties do not support an element-resolved abundance inference. A globally sub-solar APEC abundance is not an element-by-element first-ionization-potential (FIP) measurement. Temperature structure, absorption, and the fitted emission measure are covariant with abundance in these moderate-resolution spectra.

Coronal abundance differences are well documented in active stars, including Algol \citep{drak03,gud04,lam15}. However, the supplied materials contain no photospheric or chromospheric abundance measurement of the TV~Cas secondary. We therefore cannot establish its coronal-to-photospheric depletion factor or identify a FIP or inverse-FIP pattern. Such a comparison would require independent stellar abundance measurements and element-resolved X-ray spectroscopy.

\subsection{Variability and the folded light curve} \label{sec:var}

The different mean rates and within-observation variability undermine the assumption that all epochs sample a stable orbital waveform. Normalizing the constructed series illustrates sensitivity to epoch-dependent levels, but does not separate intrinsic luminosity changes from instrumental effects. Phase-confounded short observations also cannot independently determine an orbital signal and an epoch-specific normalization. The single long 2012 observation provides the clearest qualitative eclipse test; repeating such coverage would improve the distinction between orbital effects and stochastic variability.

Covariance among absorption, temperature, and abundance limits
the interpretation of differences in the fitted absorbing column.
Instrument-specific count rates alone cannot establish an
anti-correlation between absorption and intrinsic luminosity.

\section{Summary} \label{sec:summary}

We have organized the archival X-ray characterization of TV~Cas to distinguish directly supported spectral and temporal behavior from interpretations requiring additional calibration or analysis.
\begin{enumerate}
\item The six ACIS spectra are described by thermal plasma models, generally near 1~keV, with a two-temperature description for ObsID 07082. Differences in the abundance scales and confidence conventions
of the available fits limit quantitative comparisons between
observations.
\item ObsID 08458 contains the prominent flare-like event. The ACIS mean count-rate contrast is approximately 3.7, but this is not a response-corrected luminosity ratio.
\item The 2012 EPIC observation covers approximately 82\% of an orbit and shows no conspicuous deep eclipse at the predicted secondary conjunction. Intrinsic variability
and orbital-phase uncertainty limit quantitative constraints on
the eclipse depth and coronal extent.

\item The multi-epoch folded light curve shows apparent troughs.
Per-epoch normalization reduces the relative peak-to-trough
amplitude of the descriptive harmonic fit from approximately
38\% to 11\%, indicating that differences between observations
strongly influence the folded profile.

\item Two-temperature models improve the description of both
EPIC-pn spectra, with fitted temperatures of approximately
0.74 and 1.39~keV in 2003, and 0.87 and 2.16~keV in 2012.
At 240~pc, the inferred intrinsic 0.3--8~keV luminosities are
approximately $7.1\times10^{30}$ and $1.2\times10^{31}$~erg~s$^{-1}$,
respectively. Residual structure in the 2012 spectrum limits
the interpretation of its fitted parameters.

\item The thermal spectra and flare-like variability favor a
magnetically active corona on the cool secondary. Contributions
from accretion or an unresolved tertiary remain possible,
but their relative contributions are not constrained.
\end{enumerate}

\begin{acknowledgments}
We thank the anonymous referee for a constructive report that improved this paper. This research has made use of data obtained from the \textit{Chandra} Data Archive and observations obtained with \textit{XMM-Newton}, an ESA science mission with instruments and contributions directly funded by ESA Member States and NASA. We acknowledge the use of software provided by the \textit{Chandra} X-ray Center (CXC) in the application packages \texttt{CIAO} and \texttt{Sherpa}, and of \texttt{SAS}, developed and maintained by the \textit{XMM-Newton} Science Operations Centre. This work has made use of data from the European Space Agency mission \textit{Gaia}, the SIMBAD database operated at CDS, Strasbourg, France, and NASA's Astrophysics Data System. 
The author gratefully acknowledges generous philanthropic support.
\end{acknowledgments}

\software{ISIS (\url{https://space.mit.edu/cxc/isis/}), CIAO \citep{ciao}, SAS \citep{sas}, XSPEC \citep{arnaud96}, Astropy \citep{astropy13, astropy22}, NumPy \citep{numpy}, SciPy \citep{scipy}, Matplotlib \citep{matplotlib}}

\facilities{CXO (ACIS), \textit{XMM-Newton} (EPIC), Gaia}

\end{document}